\documentclass[aps,prb,showpacs,twocolumn,amssymb,floats,epsfig]{revtex4}
\usepackage{epsfig,psfrag,subfigure}
\usepackage{graphicx,psfrag,subfigure}
\usepackage{color}
\usepackage{amssymb,amsbsy,amsmath}

\newcommand\beq{\begin{equation}}
\newcommand\eeq{\end{equation}}
\newcommand\bea{\begin{eqnarray}}
\newcommand\eea{\end{eqnarray}}
\newcommand\al{\alpha}
\newcommand\be{\beta}
\newcommand\ga{\gamma}
\newcommand\de{\delta}
\newcommand\ep{\epsilon}
\newcommand\De{\Delta}
\newcommand\si{\sigma}

\newcommand\la{\lambda}
\newcommand\om{\omega}
\newcommand\ta{\theta}
\newcommand\dg{\dagger}

\newcommand\non{\nonumber}
\newcommand\noi{\noindent}

\newcommand\ig{\includegraphics}
\newcommand\bib{\bibitem}

\begin{document}

\title{Majorana edge modes in the Kitaev model}

\author{Manisha Thakurathi$^1$, K. Sengupta$^2$, and Diptiman Sen$^1$}
\affiliation{\small{
$^1$Centre for High Energy Physics, Indian Institute of Science, Bangalore
560 012, India \\
$^2$Theoretical Physics Department, Indian Association for the Cultivation
of Science, Jadavpur, Kolkata 700 032, India}}

\date{\today}

\begin{abstract}

We study the Majorana modes, both equilibrium and Floquet, which can
appear at the edges of the Kitaev model on the honeycomb lattice. We
first present the analytical solutions known for the equilibrium Majorana
edge modes for both zigzag and armchair edges of a semi-infinite Kitaev 
model and chart the parameter regimes of the model in which they appear.
We then examine how edge modes can be generated if the Kitaev coupling on 
the bonds perpendicular to the edge is varied periodically in time as
periodic $\de$-function kicks. We derive a general condition for the 
appearance and disappearance of the Floquet edge modes as a function of the 
drive frequency for a generic $d$-dimensional integrable system. We confirm 
this general condition for the Kitaev model with a finite width by mapping 
it to a one-dimensional model. Our numerical and analytical study of this 
problem shows that Floquet Majorana modes can appear on some edges in the 
kicked system even when the corresponding equilibrium Hamiltonian has no 
Majorana mode solutions on those edges. We support our analytical studies 
by numerics for finite sized system which show that periodic kicks can 
generate modes at the edges and the corners of the lattice.

\end{abstract}

\maketitle

\section{Introduction}
There have been extensive theoretical and experimental studies of
topological phases of matter in recent years~\cite{hasan,qi,nayak}.
Systems in these phases exhibit a bulk-boundary correspondence, namely, 
non-trivial topological properties of the gapped states in the bulk
are related to gapless states at the boundary.
The number of species of gapless boundary modes is typically determined 
by bulk topological invariant(s) whose nature depends on the spatial 
dimensionality of the system and its symmetries. Examples of systems with 
topological phases include two- and three-dimensional (2D and 3D) 
topological insulators, quantum Hall systems, 1D semiconducting wires 
with strong spin-orbit coupling and induced superconductivity, and
unconventional superconductors.

The Kitaev model on a honeycomb lattice and the Kitaev chain (which is 
one of the models used to describe a wire with $p$-wave superconductivity) 
provide well-known examples of systems with such 
a bulk-boundary 
correspondence~\cite{kitaev06,chen07,schmidt07,kells08,kells10}. The
physics of the bulk of these systems has a natural description in
terms of Majorana fermions. In addition, the edge physics of the
Kitaev chain and its several variants have been studied in great
detail. It is well-known that a finite length chain has Majorana
modes at its two
ends~\cite{kitaev,beenakker,lutchyn1,oreg,fidkowski2,potter,fulga,stanescu1,
tewari,gibertini,lim,tezuka,egger,ganga,sela,lobos,lutchyn2,cook,pedrocchi,
sticlet,jose,klinovaja,alicea,stanescu2,halfmetal,brouwer1,shivamoggi,
adagideli,sau,akhmerov,gott1,gott2,niu,sau2,lang,brouwer3,cai}. A number of
experimental realizations of such models have found evidence for such Majorana
modes~\cite{kouwenhoven,deng,rokhinson,das,finck}. However, the
edge states of the Kitaev honeycomb model has not been studied in as
much detail. Some discussion appears in
Refs.~[\onlinecite{kitaev06,kells10}] in the context of such states;
however the analysis of Ref.\ \onlinecite{kells10} does not address
the geometry of the edge and the full parameter range of the model.
There has also been some discussion of localized Majorana modes in
the bulk of this model in the presence of dislocations~\cite{petrova}.

Recently, there have been several studies of systems in which the Hamiltonian 
varies with time in a periodic way which gives rise to 
edge or boundary states~\cite{kita1,lind1,jiang,gu,kita2,lind2,
morell,trif,russo,basti1,liu,tong,cayssol,rudner,basti2,tomka,gomez,dora,
katan,kundu,basti3,schmidt,reynoso,wu,perez,thakurathi,reichl}. Photonic 
systems with 
edge states have been demonstrated experimentally~\cite{recht}. 
Some of the theoretical papers have studied the boundary modes 
(called Floquet modes) in these systems and the associated topological
invariants~\cite{kita1,lind1,jiang,trif,liu,tong,cayssol,rudner,kundu,
thakurathi}. In particular, Refs.~[\onlinecite{jiang,tong,thakurathi}]
have discussed Floquet modes of the Majorana type at the ends of
one-dimensional (1D) systems like the Kitaev chain. Floquet edge modes of the
Kitaev honeycomb model have, however, not been studied so far to the best of
our knowledge.

In this paper, we study the edge modes of the Kitaev honeycomb model
for both a time-independent Hamiltonian and for a periodic driving
of one of the parameters in the Hamiltonian. We consider a
semi-infinite Kitaev model and review the known analytical solutions
for the edge problem for both the zigzag and the armchair edges of
the model. The existence of edge states are known to 
depend both on the type of edge (armchair or zigzag) and on the
values of the coupling parameters of the model leading to an phase
diagram showing the presence/absence of these states. 
For any set of values of the coupling parameters of the model,
there exists a range of values of the transverse momentum $k$ for
which the edge states exist. We show that the edge modes have zero energy 
and the associated operators are of the Majorana type, corresponding to 
equal superpositions of $\pm k$ states. We also discuss the properties of
these edge states which distinguishes them from their bulk
counterparts. Our equilibrium analysis is followed by a discussion
of the formalism for studying generation of non-equilibrium Floquet
edge states in the presence of a periodic $\de$-function kick
which changes the Kitaev coupling on the bonds perpendicular to the
edge. We provide a concrete numerical method for the detection of
such Floquet edge modes via computation of the inverse participation
ratio of the eigenstates of the Floquet Hamiltonian. We also develop an
analytical understanding for the appearance and disappearance of
these Floquet edge modes as a function of the drive frequency by
providing a general formula for the momentum-dependent drive
frequency at which such phenomenon occurs for an arbitrary
$d$-dimensional integrable model. We show that the $\de$-function
kicks can generate modes on certain edges even in the parameter
regime where the time-independent Kitaev Hamiltonian has no edge
solution. For a system with infinitely long edges but finite width,
the problem can be mapped to a finite system in one dimension
running in the direction transverse to the edges; the parameters of
this 1D system depend on the couplings, the drive frequency and the
transverse momentum $k$. This reduction to one dimension enables us
to use some results from Ref.~\onlinecite{thakurathi} regarding the
Floquet Majorana modes. For a system which is finite in both
directions, we study the problem numerically and demonstrate the
existence of a variety of Floquet modes; some of these modes lie on
the edges while the others lie only at the corners of the system.

The plan of the rest of this paper is as follows. In Sec.\
\ref{basics}, we review some of the properties of the Kitaev
honeycomb model, its energy-momentum dispersion in the bulk, and the
phase diagram. This is followed by Sec.\ \ref{edgeeq}, where we
study the edge modes in equilibrium. In Sec.\ \ref{edgefl1}, we
provide a discussion of the formalism for detection of the Floquet
edge modes. This is followed by Sec.\ \ref{edgefl2} where we apply
this formalism to the Kitaev model with periodic $\de$-function kicks. 
We show how the problem can be mapped to a one-dimensional system thus 
enabling us to analytically find the driving frequencies where edge 
modes appear or disappear. Finally, we conclude in Sec.\ \ref{diss}.

\section{Kitaev Honeycomb Model}
\label{basics}

The Kitaev model consists of spin-$1/2$'s placed on the sites of a
honeycomb lattice with a Hamiltonian of the form \beq
H~=~\sum_{j+l=even}^{} (J_1 \si^x_{j,l} \si^x_{j+1,l} + J_2
\si^y_{j-1,l} \si^y_{j,l} + J_3 \si^z_{j,l} \si^z_{j,l+1}),
\label{hsig}\eeq where $j,l$ are the column and row indices
respectively, $\si^a_{m,n}$ are Pauli matrices at the site labeled
$(m,n)$, and $J_1, ~J_2$ and $J_3$ are the coupling parameters. In
this section we will assume that all the couplings are
time-independent. Let us also assume that all the $J_i \ge 0$.

\begin{figure}[htb] \ig[width=2.8in]{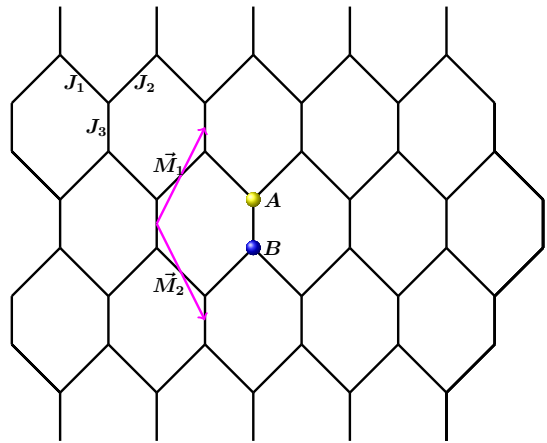}
\caption[]{(Color online) Kitaev model on the honeycomb lattice with $xx$
coupling $J_1$, $yy$ coupling $J_2$ and $zz$ coupling $J_3$. $\vec{M_1}$
and $\vec{M_2}$ are the spanning vectors of the lattice, and $A$ and $B$
denote the two sites of a unit cell.} \label{fig01} \end{figure}

A picture of the honeycomb lattice is shown in Fig.~\ref{fig01}. We take the
unit cells of the lattice to be the vertical bonds with sites labeled $A$
and $B$; these have $j+l$ equal to odd and even integers respectively.
If the number of sites is denoted by $N$ (assumed to be even), the
number of unit cells is $N/2$. It is convenient to set the nearest-neighbor
distance to be $1/\sqrt{3}$. Each unit cell is then labeled by a vector
$\vec{n}= \hat{i} n_1+(\frac{1}{2} \hat{i} + \frac{\sqrt{3}}{2}
\hat{j}) n_2$, where $n_1, ~n_2$ are integers which are related to the
coordinates of the $B$ site in that unit cell as $n_1 = (j-l)/2$ and
$n_2 = l$. Fig.~\ref{fig01} shows
the spanning vectors $\vec{M_1} =\frac{1}{2} \hat{i} + \frac{\sqrt{3}}{2}
\hat{j}$ and $\vec{M_2} =\frac{1}{2} \hat{i} - \frac{\sqrt{3}}{2} \hat{j}$
which join some neighboring unit cells.

We now introduce the Majorana operators~\cite{kitaev06,chen07}
\bea \hat a_{j,l} &=& \left( \prod_{i=-\infty}^{j-1} \si^z_{i,l} \right)
\si^y_{j,l},~~ \text{for} ~j+l=\text{even} \non \\
\hat b_{j,l} &=& \left( \prod_{i=-\infty}^{j-1} \si^z_{i,l} \right)
\si^x_{j,l}, ~~ \text{for} ~j+l=\text{odd}. \eea
These are Hermitian operators satisfying the anticommutation relations
$\{ \hat a_{m,n}, \hat a_{m',n'} \} = 2 \de_{mm'} \de_{nn'}$, $\{ \hat b_{m,n},
\hat b_{m',n'} \} = 2 \de_{mm'} \de_{nn'}$, and $\{ \hat a_{m,n}, 
\hat b_{m',n'} \} = 0$. In terms of these operators the Hamiltonian takes the 
form
\beq H~=~ i~\sum_{\vec{n}}^{} (J_1 \hat b_{\vec{n}} \hat a_{\vec{n}-\vec{M_1}}
+ J_2 \hat b_{\vec{n}} \hat a_{\vec{n}-\vec{M_1}} + J_3 \hat D_{\vec{n}} 
\hat b_{\vec{n}} \hat a_{\vec{n}}). \label{hmaj} \eeq
The $\hat D_{\vec{n}}$'s are operators which commute with each other and 
with the Hamiltonian; their eigenvalues can take the
values $\pm 1$ independently for each $\vec{n}$, thereby decomposing the
$2^N$-dimensional Hilbert space into $2^{N/2}$ sectors. It is known that the
ground state of the model lies in the sector in which $\hat D_{\vec{n}} = 1$ 
for all $\vec{n}$; we will work in this sector throughout this paper.

The Fourier transforms of the Majorana operators are defined as
\bea \hat a_{\vec{n}} &=&\sqrt{\frac{4}{N}} \sum_{\vec{k}\in \frac{1}{2}
\text{BZ}}{}(\hat a_{\vec{k}} e^{i\vec{k} \cdot \vec{n}}+ \hat a_{\vec{k}}^\dg
e^{-i\vec{k} \cdot \vec{n}}), \non \\
\hat b_{\vec{n}} &=&\sqrt{\frac{4}{N}} \sum_{\vec{k}\in \frac{1}{2}
\text{BZ}}{}(\hat b_{\vec{k}} e^{i\vec{k} \cdot \vec{n}}+ \hat b_{\vec{k}}^\dg
e^{-i\vec{k} \cdot \vec{n}}), \label{fourier} \eea
which satisfy the anticommutation relations $\{ \hat a_{\vec{k}}, 
\hat a_{\vec{k}'}^\dag \} = \{ \hat b_{\vec{k}}, \hat b_{\vec{k}'}^\dag \} = 
\de_{\vec{k},\vec{k}'}$. Note that the sums over $\vec k$ in 
Eq.~\eqref{fourier} only go over half the Brillouin zone (BZ);
a convenient choice of the BZ is given by a rhombus whose vertices lie
at $(k_x,k_y)= (\pm 2\pi, 0)$ and $(0,\pm 2\pi /\sqrt{3})$.
The Hamiltonian in Eq.~\eqref{hmaj} can then be written in momentum space as
\bea H &=& \sum_{ \vec{k}\in \frac{1}{2}\text{BZ}} ~\left( \begin{array}{cc}
\hat a_{\vec{k}}^\dg & \hat b_{\vec{k}}^\dg \end{array} \right) ~H_k ~\left(
\begin{array}{c}
\hat a_{\vec{k}} \\
\hat b_{\vec{k}} \end{array} \right), \non \\
H_{\vec{k}} &=& 2 [J_1 \sin(\vec{k}.\vec{M}_1)-J_2 \sin(\vec{k}.\vec{M}_2)]
\tau^x \non \\
&& + 2 [J_3+J_1 \cos(\vec{k}.\vec{M}_1)+J_2 \cos(\vec{k}.\vec{M}_2)] \tau^y,
\label{hmom} \eea
where $\tau^a$ are Pauli matrices denoting pseudospin. The dispersion relation
can be derived from Eq.~\eqref{hmom}; it consists of two bands with energies
\bea E_{\vec{k}}^\pm &=& \pm 2[\{J_1 \sin(\vec{k}.\vec{M}_1)-J_2 \sin(\vec{k}.
\vec{M}_2)\}^2 \non \\
&& +\{J_3+J_1 \cos(\vec{k}.\vec{M}_1)+J_2 \cos(\vec{k}.\vec{M}_2)\}^2 ]^{1/2}.
\label{disp1} \eea

The phase diagram of the model can be deduced from Eq.~\eqref{disp1}. Given
that $J_i \ge 0$, it is convenient to normalize them so that $J_1 + J_2 + J_3
= 1$. This describes points lying within (or on) an equilateral triangle. This
triangle can be divided into four smaller equilateral triangles as shown in
Fig.~\ref{fig03}, namely, $A_x$ where $J_1 > J_2 + J_3$, $A_y$ where $J_2 >
J_1 + J_3$, $A_z$ where $J_3 > J_1 + J_2$, and $B$ where each of the $J_i$ is
less than the sum of the other two. It turns out~\cite{kitaev06} that the
system is gapped in the three $A$ phases, with $E_{\vec{k}}$ being non-zero
for all $\vec k$, and is gapless in the $B$ phase, with $E_{\vec{k}} = 0$ for
some value of $\vec k$ whose value depends on the location of the point in
that phase. The four phases are separated from each other by quantum critical
lines where one of the couplings is equal to the sum of the other two.

\section{Phase diagram for edge states}
\label{edgeeq}

In this section,
we will consider two kinds of edges for the honeycomb lattice, namely, zigzag
and armchair~\cite{nakada,kohmoto}. These are shown in Figs. 2 and 4
respectively. We will assume that the edges are infinitely long; translational
invariance then implies that the edge states can be labeled by their momentum
$k$. We will analytically study the ranges of the couplings $J_i$ for which
edge states exist for these two kinds of edges. (In principle there can be
more complicated kinds of edges, but analytical results for the edge states
are then no longer available).

To find the edge states, we first write the Hamiltonian in
Eq.~\eqref{hmaj} in the form
\beq H ~=~ 2i ~\sum_{\al \be} ~\hat b_\be ~L_{\be \al} ~\hat a_\al, 
\label{hgen} \eeq
where $\al,~\be$ label the sites, and $L_{\al \be}$ is a real matrix.
We now use the Heisenberg equations of motion $d \hat a_\al / dt ~=~ i ~[H,
\hat a_\al]$ and similarly for $\hat b_\be$. We then obtain
\bea \frac{d\hat a_\al}{dt} &=& - 4~ \sum_{\be} ~\hat b_\be ~L_{\be \al}, 
\non \\
\frac{d\hat b_\be}{dt} &=& 4~ \sum_{\al} ~L_{\be \al} ~\hat a_\al. 
\label{eom1} \eea

We note that the Hamiltonian in Eq.~\eqref{hgen} and the time evolution
given in Eqs.~\eqref{eom1} are invariant under an effective time-reversal 
transformation which complex conjugates all numbers, and takes $t \to -t$, 
$\hat a_\al \to \hat a_\al$ and $\hat b_\be \to - \hat 
b_\be$~\cite{thakurathi,gottardi}. Such a symmetry ensures that all the zero 
energy modes (to be discussed below) involve only the $\hat a$ operators or 
only the $\hat b$ operators, not combinations of the two. Thus all the edge 
states, in contrast to their bulk counterparts, have weights on either 
$A$ or $B$ sublattices of the honeycomb, but not both.

We will now see that for appropriate ranges of couplings, there are states
which have zero energy and are localized near a particular edge. We note that
our analysis is similar to that used to find edge states in
graphene~\cite{nakada,kohmoto} and other systems~\cite{cortes,dutreix}, 
except that we are considering Majorana fermions rather than ordinary
fermions~\cite{klinovaja2}. We will find the wave functions of these states
by solving Eqs.~\eqref{eom1}. We will henceforth denote wave functions by 
alphabets without hats (such as $a$ and $b$) to distinguish them from 
operators which are denoted by $\hat a$ and $\hat b$.

\subsection{Zigzag Edge}
\begin{figure}[htb] \ig[width=2.8in]{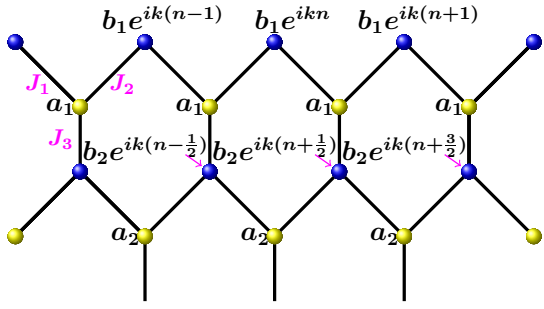}
\caption[]{(Color online) Zigzag edge. Majorana fermions with a momentum $k$
along the edge are indicated.} \label{fig02} \end{figure}

We look for a state with momentum $k$ at the zigzag edge at the top of a
system as shown in Fig.~\ref{fig02}; $k$ lies in the range $-\pi$ to $\pi$.
In that figure, the wave functions for the Majorana operators of type $\hat b$
are given by $b_{m,n}$, where $n$ goes from $-\infty$ to $\infty$ and increases
towards the right along the edge, and $m=1,2,3,...$ increases as we go down
away from the edge and into the bulk of the system. Further, we will take
$b_{m,n} = b_m e^{ikn}$ or $b_m e^{ik(n+1/2)}$ depending on whether $m$ is 
odd or even. Similarly, the wave functions for Majorana operators of type
$\hat a$ are given by $a_{m,n} = a_m e^{ikn}$ or $a_m e^{ik(n+1/2)}$; these
factors are not shown in Fig.~\ref{fig02}.

\begin{figure}[htb] \ig[width=2.2in]{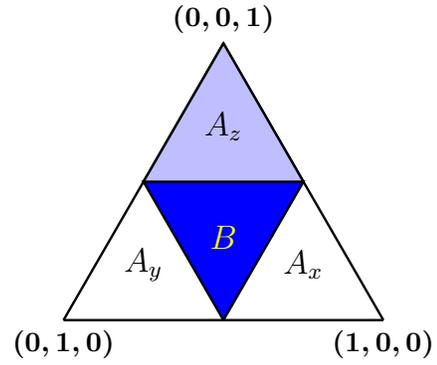}
\caption[]{(Color online) Phase diagram for zigzag edge states in the triangle
with $J_1+J_2+J_3=1$. Majorana modes exist in the regions $A_z$ and $B$.}
\label{fig03} \end{figure}

We then discover that the Heisenberg equations of motion in Eqs.~\eqref{eom1}
have {\it zero energy} solutions (i.e., with $d \hat a_\al/dt=0$ and 
$d \hat b_\be /dt = 0$) in which $a_m = 0$ for all $m$, and
\beq J_1 b_m e^{ikn}+ J_2 b_m e^{ik(n+1)}+ J_3 b_{m+1} e^{ik(n+\frac{1}{2})}
=0 \eeq
for all $m \ge 1$. This is solved by assuming that $b_m=(\la_k)^m$; we then 
get 
\beq \la_k = -\frac{J_1 e^{-ik/2}+ J_2 e^{ik/2}}{J_3}. \label{zig}\eeq
For a normalizable edge state, we require $|\la_k| < 1$. According to
Eq.~\eqref{zig} this occurs if
\beq \cos k ~<~ \frac{J_3^2-J_1^2-J_2^2}{2 J_1 J_2}. \label{zigcon} \eeq
We discover that Eq.~\eqref{zigcon} is valid for all values of $k$ in region
$A_z$ and for a finite range of values of $k$ in region $B$. Since 
Eq.~\eqref{zigcon} has a solution with $-k$ if it has a solution with $+k$, 
and $\la_{-k} = \la_k^*$ according to Eq.~\eqref{zig}, we can superpose these 
two wave functions (along with appropriate creation and annihilation 
operators) to obtain a Hermitian solution for $\hat b_{m,n}$ of the form 
$e^{ikn} (\la_k)^m \hat b_k+ e^{-ikn} (\la_{-k})^m \hat b_k^\dg$.
(For $k=0$ we directly get a real wave function of the form $\la_0^m$, where 
$\la_0 = - (J_1 + J_2)/J_3$, and therefore a Hermitian solution of the form 
$\la_0^m \hat b_0$; such a state exists everywhere in region $A_z$). 
In regions $A_x$ and
$A_y$, Eq.~\eqref{zigcon} is not satisfied for any value of $k$; hence there
are no zigzag edge states in these two regions. Fig.~\ref{fig03} shows the 
phase diagram where Majorana states of type $\hat b$ exist at a zigzag edge
at the top edge of the system. The length scale over which an edge state
decays into the bulk is given by $\xi_k = - 1/\ln | \la_k|$.


A similar analysis shows that the zigzag edge at the bottom of the Kitaev
system will have Majorana states of type $\hat a$ (i.e., with $b_m = 0$ for all
$m$) in the same regions as shown in Fig.~\ref{fig03}. These statements assume
that the top and bottom edge are separated by a distance which is much larger 
than the decay length $\xi_k$. If the separation between the edges is 
comparable to $\xi_k$ for some value of $k$, the two edge states will 
hybridize to give two states with energies different from zero.

\subsection{Armchair Edge}
\begin{figure}[htb] \ig[width=2.8in]{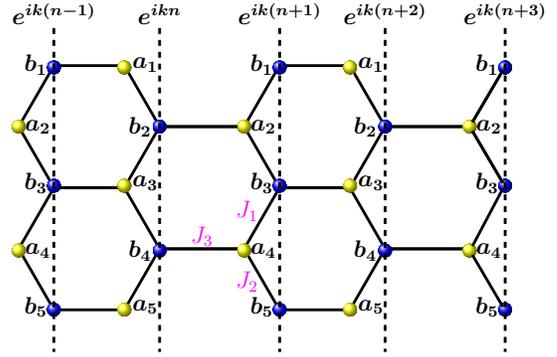}
\caption[]{(Color online) Armchair edge. Majorana fermions with a momentum $k$
along the edge are indicated.} \label{fig04} \end{figure}

We now look for a state at an armchair edge with momentum $k$ as shown in
Fig.~\ref{fig04}. The $\hat a$ and $\hat b$ Majorana operators have 
wave functions given by $a_m
e^{ikn}$ and $b_m e^{ikn}$; the figure shows these factors only for $b_m$.
Note that this figure is obtained by rotating Fig.~\ref{fig02} by $\pi/2$
so that the horizontal bonds have couplings $J_3$ in Fig.~\ref{fig04}.

\begin{figure}[htb] \ig[width=2.2in]{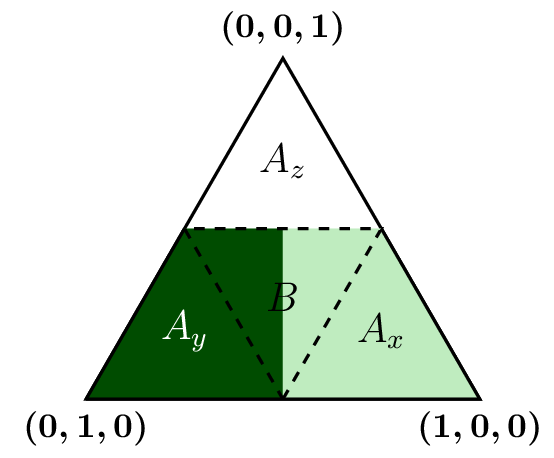}
\caption[]{(Color online) Phase diagram for armchair edge states in the
triangle with $J_1+J_2+J_3=1$. Majorana modes of type $\hat b$ ($\hat a$) 
exist in the regions $A_y$ ($A_x$) and the left (right) half of $B$. These 
regions are indicated by dark (light) shades.} \label{fig05} \end{figure}

We find that the Heisenberg equations of motion have zero energy
solutions with $a_m = 0$ for all $m$, provided that \beq J_1 b_m +
J_2 b_{m+2} + J_3 b_{m+1} e^{-ik} ~=~ 0 \label{arm1} \eeq for all $m
\ge 1$, and \beq J_2 b_2 + J_3 b_1 e^{-ik} ~=~ 0. \label{arm2} \eeq
Assuming $b_m=(\la_k)^m$, we get from Eq.~\eqref{arm1} \beq \la_k^2
+ \frac{J_3}{J_2} e^{-ik} \la_k +\frac{J_1}{J_2}=0. \label{armcon}
\eeq This has the solutions \beq \la_{k\pm} ~=~ \frac{1}{2J_2}
\left[ - ~J_3 e^{-ik} ~\pm~ \sqrt{J_3^2 e^{-2ik} ~-~ 4
J_1J_2}\right]. \eeq These two roots satisfy the equations \beq
\la_{k+} ~+~ \la_{k-} ~=~ -~ (J_3/J_2) ~e^{-ik}, \label{eig1}\eeq
\beq \la_{k+} \la_{k-} ~=~ J_1/J_2. \label{eig2}\eeq
Eqs.~(\ref{arm1}-\ref{arm2}) imply that a normalizable edge state
will exist if both $|\la_{k\pm}| < 1$. Eq.~\eqref{eig2} then implies
that we must have $J_1 < J_2$. This condition and $|\la_{k\pm}| < 1$
together imply that $|\la_{k+} + \la_{k-}| \le 1 + (J_1/J_2)$.
Substituting this in Eq.~\eqref{eig1}, we obtain the condition $J_3
\le J_1 + J_2$. Putting these together with $J_1 < J_2$, we obtain
the dark shaded region on the left side of Fig.~\ref{fig05} where
zero energy edge modes of type $\hat b$ exist for some values of $k$.
Combining states with $\pm k$ will again give us Majorana operators which
are Hermitian. A similar analysis shows that zero energy edge
modes of type $\hat a$ (i.e., with $b_m = 0$ for all $m$) exist in the
light shaded region on the right side of Fig.~\ref{fig05}, namely,
in the region with $J_2 < J_1$ and $J_3 \le J_1 + J_2$.

Before ending this subsection, we note that the solution we have found 
for a zigzag edge state is an extension of the one in standard graphene 
with isotropic hoppings \cite{nakada}; the complex version of our solution 
corresponding to a single value of the momentum $k$ (rather than a Hermitian
superposition of $\pm k$) reproduces the graphene 
edge state for the special case $J_1=J_2=J_3$. For an armchair edge, on the 
other hand, there is no solution for $J_1=J_2=J_3$ and therefore no solution
in standard graphene. However, armchair edge states can be found for strained 
graphene with anisotropic hoppings \cite{kohmoto}.

\subsection{Finite Systems}
In Secs. III A and III B, we considered systems whose edges are infinitely
long and are therefore translationally invariant. This allowed us to
effectively map the system to a 1D problem which is
characterized by the parameters $J_i$ and the edge momentum $k$.

\begin{figure}[htb] \ig[width=2.8in]{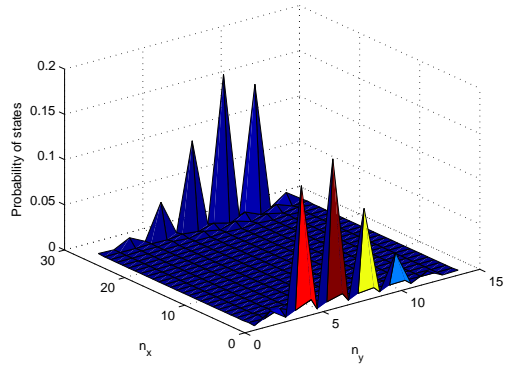}
\caption[]{(Color online) Armchair edge states for a system with $N_x \times
N_y = 27 \times 14$ sites, with $J_1=0.7, ~J_2=0.15$ and $J_3=0.15$.}
\label{fig06} \end{figure}

\begin{figure}[htb] \ig[width=2.8in]{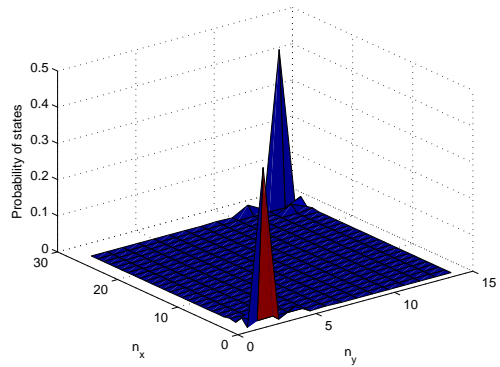}
\caption[]{(Color online) Corner states for a system with $N_x \times N_y
= 27 \times 14$ sites, with $J_1=0.7, ~J_2=0.15$ and $J_3=0.15$.}
\label{fig07} \end{figure}

In this section, we will numerically study finite systems which are not
translationally invariant. We consider a system which has zigzag edges
along the $x$ direction and armchair edges along the $y$ direction.
Specifically, we consider a system with $N_x \times N_y = 27 \times 14$ sites,
with $J_1=0.7, ~J_2=0.15$ and $J_3=0.15$; this lies in the region $A_x$ in the
phase diagram in Fig.~\ref{fig05}. For these parameter values, the discussion
in the previous subsection implies that there should be edge modes on the
armchair edges. These will not be at exactly zero energy due to hybridization
between the two armchair edges on the opposite sides of the system. However,
we find that their energies are quite close to zero since the distance
between the two edges is $N_x = 27$ is much larger than the lattice spacing;
we will therefore continue to call them Majorana modes. Numerically we find a
total of 14 Majorana edge modes. This agrees with what we expect from
Fig.~\ref{fig04}; on each of the armchair edges, the number of Majorana modes
should be equal to the number of either $a$ or $b$ sites, and this number is
equal to $N_y/2 = 7$. Interestingly, if we look at the wave functions of all
the edge modes, we find that 12 of them are localized along the armchair edges
as expected (see Fig.~\ref{fig06} for an example), but the remaining 2 are
localized at the corners as shown in Fig.~\ref{fig07}.

Let us now consider the same $27 \times 14$
system but change the couplings to $J_1 = 1/6$,
$J_2 = 1/6$ and $J_3 = 2/3$. According to Fig.~\ref{fig03}, this lies in the
$A_z$ phase and should therefore only have states on the zigzag edges. Indeed
we find numerically that there are 26 edge states; of these 13 are at the top
zigzag edge and 13 are at the bottom zigzag edge. This is expected since the
top (bottom) row with 27 sites has 13 sites of type $\hat b$ ($\hat a$).

\subsection{Properties of the edge states}

From the numerical studies of the previous subsection and from
general analytical results obtained in earlier subsections, we have
confirmed that the number of edge modes is exactly half the number
of sites at the edge for an armchair edge. This fact is reminiscent
of the edge states at the ends of unconventional superconductors for
which one finds exactly half the number of states as the number of
transverse momentum modes \cite{sengupta1}. In this section, we
explore this property a little further. To this end, let us consider
a zigzag edge and define a two-component fermion for a semi-infinite
2D Kitaev model $\hat \psi^\dg_{m,k} = [\hat a^\dg_{m,k}, \hat b^\dg_{m,k}]$, 
where $k$ denotes the momentum along the edge and $m$ denotes the coordinate
in the direction perpendicular to the edge (see Fig.~\ref{fig02}). (The 
operators $\hat a^\dg_{m,k}$ and $\hat b^\dg_{m,k}$ are obtained by Fourier
transforming $\hat a_{m,n}$ and $\hat b_{m,n}$ as explained below). We then
define a correlation matrix $C$ whose elements are given by
\bea C_{11} &=& \sum_m ~\langle \hat a^\dg_{m,k} \hat a_{m,k} + 
\hat b^\dg_{m,k} \hat b_{m,k} \rangle, \non \\
C_{12}&=& C_{21} ~=~ \sum_m ~\langle i\hat a^\dg_{m,k} \hat b_{m,k} - i
\hat b^\dg_{m,k} \hat a_{m,k} \rangle, \non \\
C_{22} &=& \sum_m ~\langle \hat a^\dg_{m,k} \hat a_{m,k} - \hat b^\dg_{m,k} 
\hat b_{m,k} \rangle, \label{cormat} \eea
where $\langle \cdots \rangle$ implies properly normalized sums over $m$ and
is taken with respect to a state with a fixed energy and momentum $k$.

Let us evaluate the matrix $C$ for the bulk states in the limit of large 
$J_3$. In this limit the diagonal elements can be shown to be zero (to show 
this we have to ignore a constant which comes from on-site terms like 
$\hat a_{m,n}^2 = \hat b_{m,n}^2 = 1$, as explained after Eq.~\eqref{b1ktot}),
while the off-diagonal elements give $\pm 1$ if the state is
occupied; hence the eigenvalues $\lambda_i$ of $C$ are $\pm 1$. In
contrast, for a Majorana mode localized at one of the edges (on, say, the
$A$ sublattice), the off-diagonal components are zero while the
diagonal components yield $1$ so that $C$ has doubly degenerate
eigenvalue $\lambda_1=\lambda_2=1$ for these states. For a Majorana mode
localized at the other edge on the $B$ sublattice, the diagonal components
and hence the eigenvalues are $\pm 1$. Thus, for all single Majorana occupied
states, edge or bulk, the eigenvalues of $C$ assume integer values. These
results can be easily extended for all values of the couplings $J_i$.

Let us now consider a situation where a Kitaev system is in a gapped
phase with localized zero energy edge states present at the zigzag
edges, on one sublattice at the top edge and the other sublattice on
the bottom edge; the two edges are assumed to be very far from each
other. Let us consider tunneling a bulk Majorana fermion from
another Kitaev system (which is gapless) with zero energy and wave
function $(u,v)= (1,1) \exp [i (k_1 m + k_2 n)]/\sqrt{2}$. Since
the zero energy states of the gapped Kitaev system only reside at
the edges, the Majorana fermion must, after tunneling, divide
between the two zigzag edges. Thus the state of the Majorana
particle must have the form $|\psi\rangle = \alpha |A\rangle + \beta
|B\rangle$ where $|A\rangle$ is a wave function localized along one
edge with weight only on the $A$ sublattice and $|B\rangle$ is
localized along the other edge with weight only on the $B$
sublattice. In the absence of any perturbations which break
sublattice symmetry, we will have $|\alpha|^2=|\beta|^2=1/2$. For
this state, we will have $C_{12}=C_{21}=0$ (since the edges are far
from each other), while $C_{11}=1$ and $C_{22}=0$; this again leads
to integer eigenvalues. Thus for any Majorana state, the eigenvalues
of $C$ will always be a positive or negative integer; for an
unoccupied Majorana state, $C=0$ by definition. Thus the behavior of the
$|\lambda_i|$ is analogous to the properties of the expectation value of
the number operator for fermions. The fluctuations to this
expectation value can also be calculated and shown to vanish.

Next we consider a local correlation where the sum over $m$ for the
elements of $C$ is taken over a finite number of lattice sites
starting from a given edge; we choose the finite number to be much
larger than the decay lengths of all the Majorana modes localized at
that edge. Let us define the corresponding matrix as $C'$ Then for a
split Majorana one has $C'_{11} = 1/2$, $C'_{22} = \pm 1/2$ [where the
$+(-)$ sign corresponds to the state localized at that edge having weights
on the $A(B)$ sublattice respectively], and $C'_{12}=C'_{21} =0$. This leads
to fractional eigenvalues for $C'$. Thus the edge states of the Kitaev
model provide us with a way of spatially separating the two
sublattice components of the Majorana wave function leading to
fractional expectation values for local correlation functions. One
can easily show that the number operator for fermions constructed
out of the Majorana will also have half-integer expectation value. However,
the difference between the present situation and the well-known example
of electron fractionalization found in the literature~\cite{rajaraman} (in
the context of polyacetylene and quantum field theoretic models in one
dimension) is that the fluctuations from this expectation value are not small
here. We will show below that these states have either $\langle
(\hat a_{m,k}^\dg \hat a_{m,k})^2 \rangle = 1/2$ or $\langle (\hat b_{m,k}^\dg 
\hat b_{m,k})^2 \rangle = 1/2$ depending on whether the states have weight on 
the $A$ or $B$ sublattice, and that there is a finite variance which signifies 
large fluctuations from the expectation value. Thus the fractionalization of
the expectation value does not amount to fractionalization of the
eigenvalues of the correlation matrix of the Majorana fermions.

The difference of the present situation from the standard electron
fractionalization found in the literature~\cite{rajaraman} can be
understood in a number of ways. In the limit of large $J_3$, the
Majorana modes near the top zigzag edge in Fig.~\ref{fig02} are
completely localized at the sites of the top row labeled as
$b_{1,n}$, while the Majorana modes near the bottom zigzag edge are
completely localized at the sites of the bottom row labeled as $a_{N_y,n}$, 
where $N_y$ is the width of the system assumed to be much larger than 1. 
Let us introduce the Fourier transform of the operators $b_{1,n}$ as 
\beq \hat b_{1,n} ~=~ \int_0^\pi ~\frac{dk}{2\pi} ~[~ \hat b_{1,k} e^{ikn} ~
+~ \hat b_{1,k}^\dg e^{-ikn} ~], \eeq 
The inverse of this is given by
\bea \hat b_{1,k} &=& \sum_{n=-\infty}^\infty ~\hat b_{1,n} e^{-ikn}, \non \\
\hat b_{1,k}^\dg &=& \sum_{n=-\infty}^\infty ~\hat b_{1,n} e^{ikn}. \eea
We can similarly define Fourier transforms of the operators $a_{N_y,n}$,
called $a_{N_y,k}$. Next, we have to find the ground state of the system.
Since the modes labeled by $b_{1,k}$ and $a_{N_y,k}$ have zero energy for all
values of $k$ if $J_3$ is infinitely large, the ground state has an enormous
degeneracy. To break this degeneracy, let us assume that $J_1$ and $J_2$ are
slightly different from zero. This will introduce a small tunneling between
the top and bottom rows of the form
\beq \De H ~=~ \int_0^\pi ~\frac{dk}{2\pi} ~[\ga_k \hat b_{1,k}^\dg 
\hat a_{N_y,k} ~+~ h.c.], \label{hamtun} \eeq
where $\ga_k$ is the tunneling amplitude which is exponentially small:
$\ga_k \sim e^{-N_y/\xi_k}$, where $\xi_k$ is the decay length of the mode
$k$. The Hamiltonian in Eq.~\eqref{hamtun} has a unique ground
state of the form
\beq | gs \rangle ~=~ \prod_k ~(u_k b_{1,k}^\dg ~+~ v_k a_{N_y,k}^\dg) ~|vac
\rangle, \eeq
where $|u_k|^2 = |v_k|^2 = 1/2$. We now see that at the top edge,
\beq \langle gs | \hat b_{1,k}^\dg \hat b_{1,k} | gs \rangle ~=~ \langle gs |
(\hat b_{1,k}^\dg \hat b_{1,k})^2 | gs \rangle ~=~ 1/2, \eeq
implying that the variance, $\langle gs | (\hat b_{1,k}^\dg \hat b_{1,k})^2 | 
gs \rangle - \langle gs | \hat b_{1,k}^\dg \hat b_{1,k} | gs \rangle^2 = 1/4$,
is not small.

Another difference between Majorana fermions and standard electrons
is as follows. The operator appearing in the diagonal component of
Eq.~\eqref{cormat}, restricted to the top row given by $m=1$ in
Fig.~\ref{fig02}, is given by
\beq \hat b_{1,k}^\dg \hat b_{1,k} ~=~ \sum_{n, n' = - \infty}^\infty ~
\hat b_{1,n} \hat b_{1,n'} e^{ik(n-n')}, \label{b1k} \eeq
which involves operators which are extremely non-local in space. Even if
Eq.~\eqref{b1k} is integrated over $k$, we still get a non-local expression
\beq \int_0^\pi ~\frac{dk}{2\pi} ~\hat b_{1,k}^\dg
\hat b_{1,k} ~=~ - ~\frac{2i}{\pi}~ \sum_{n = -\infty}^\infty
\sum_{r=0}^\infty ~\frac{\hat b_{1,n} \hat b_{1,n+2r+1}}{2r+1}, 
\label{b1ktot} \eeq
plus an infinite constant coming from $\hat b_{1m}^2 =1$. The non-local form
originates from the fact that $k$ is integrated over only half the Brillouin 
zone, i.e., $0 \le k \le \pi$. This, in turn, arises from the fact that the 
$\hat b_{1,k}$ are Fourier transforms of $\hat b_{1,n}$ which are Hermitian
operators, namely, the Majorana fermions are indistinguishable from their
antiparticles. The expression in Eq.~\eqref{b1ktot} is to be
contrasted with the total number operator for electrons which is
always given by a sum over operators which are local in space.

The above arguments for the fractionalization of expectation values and the
non-locality of Majorana modes at zigzag edges will hold at all points in
the phase $A_z$ in Fig.~\ref{fig03}. We have shown in Sec. III A that there
are Majorana modes near both the zigzag edges but residing entirely
on opposite sublattices, for all values of $k$ lying in the range
$[0,\pi]$. If we choose the finite number of lattices in the definition of
the local correlation $C'$ to be much larger than the decay length $\xi_k$
for all values of $k$, the eigenvalues of $C'$ will be $\pm 1/2$.

\section{Floquet Evolution}
\label{edgefl1}

We will now consider what happens when the Hamiltonian varies
periodically in time with a time period $T$~\cite{thakurathi}.
Namely, we will assume that the matrix $L$ in Eq.~\eqref{hgen}
changes with time in such a way that $L(t+T)=L(t)$.
Eqs.~\eqref{eom1} and their solution can be written as matrix
equations as follows. Given a system with $N=N_x N_y$ sites, let us
introduce a $(2N)$-dimensional column called $\hat c$ whose first $N$
entries are given by $(\hat a_1,\hat a_2,\cdots,\hat a_N)^T$ and last $N$ 
entries are given by $(\hat b_1,\hat b_2,\cdots, \hat b_N)^T$. Given the 
$N$-dimensional matrix $L$, we define a $(2N)$-dimensional real antisymmetric 
matrix $M$ by the block form 
\beq M ~=~ \left( \begin{array}{cc}
0 & - L^T \\
L & 0 \end{array} \right). \eeq
Eqs.~\eqref{eom1} can then be written as $d\hat c(t)/dt = 4 M(t) \hat c(t)$.
The periodicity of $M(t)$ in time implies that the solution of this equation
is given by
\bea \hat c(T) &=& U(T,0) ~\hat c(0), \non \\
{\rm where}~~~ U(T,0) &=& {\cal T} e^{4 \int_0^T dt M(t)}, \eea
and $\cal T$ denotes the time-ordering symbol. $U(T,0)$ is called the Floquet
operator. It is a unitary matrix (in our case it is real and orthogonal), and
it can be computed numerically for a given form of $M(t)$.

The eigenvalues of $U(T,0)$, called Floquet eigenvalues (FE), are given by
phases, $e^{i\ta_j}$, and they come in complex conjugate pairs if $e^{i\ta_j}
\ne \pm 1$. If $U(T,0)$ has eigenvalues $\pm 1$, the corresponding 
eigenvectors can be shown to be real.

In the next section, we will present our results for eigenvectors
of $U(T,0)$ which are localized near the edges of the honeycomb lattice and
whose FE are equal to $\pm 1$. In order to find these edge modes, we will 
use the same numerical methods as in Ref.~\onlinecite{thakurathi}. Namely, 
we will first use the inverse participation ratio to identify eigenvectors 
of the Floquet operator which are localized near the edges. [Given 
an eigenvector $\psi_j$, normalized so that $\sum_{m=1}^{2N} |\psi_j (m)|^2 
= 1$, we define its inverse participation ratio as $I_j = \sum_{m=1}^{2N} 
|\psi_j (m)|^4$. Eigenvectors with larger values of $I_j$ are more localized 
in space]. We will then check if these eigenvectors are real and if their FE 
($\pm 1$) are separated from all the other FE by a finite gap as the 
dimensions of the system, $N_x$ and $N_y$, are made very large. If all these 
conditions are met, these eigenvectors will be called Floquet Majorana modes.

\section{Periodic $\de$-function kicks}
\label{edgefl2}

In this section, we will study what happens when one of the
parameters in the Kitaev honeycomb model is given $\de$-function
kicks periodically in time. The reason for studying this kind of a
periodic variation is that it is easy to study both numerically and
analytically~\cite{stock}.

Let us first consider what happens if $J_3$ in Eq.~\eqref{hmaj} is
periodically kicked, so that
\beq J_3 (t) ~=~ J_0 ~+~ J_p ~\sum_{n=-\infty}^\infty ~\de (t - nT),
\label{ht1} \eeq
where the time period $T$ is related to the drive frequency as $T=2 \pi/\om$.

We numerically compute the operator $U(T,0)$
for various values of the parameters $J_1$, $J_2$, $J_0$, $J_p$, $\om$ and
the system size $N_x \times N_y$. We then find all the eigenvalues and
eigenvectors of $U(T,0)$ and use the inverse participation ratio and the
eigenvectors to identify the Floquet Majorana modes as described above.

The Floquet operator is given by a product of two exponentials
\beq U(T,0) ~=~ e^{4M_1} ~e^{4M_0 T}, \label{flo1x} \eeq
where $e^{4M_0 T}$ is the operator which time evolves from $t=0$ to $t=T$,
and $e^{4M_1}$ then evolves across the $\de$-function at $t=T$.

\begin{figure}[htb] \ig[width=2.8in]{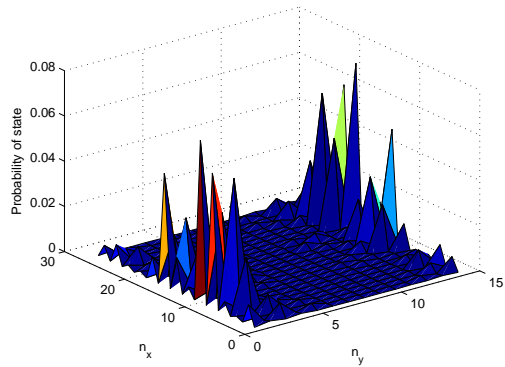}
\caption[]{(Color online) Zigzag edge states for a system with
$N_x \times N_y = 27 \times 14$, $J_1=0.7, ~J_2=0.15, ~J_0=0.15, ~J_p=0.2$
and $\om=3$.} \label{fig08} \end{figure}

To illustrate the Floquet Majorana modes, we now
consider a system with $N_x \times N_y =27 \times 14$ sites with
$J_1 =0.7, ~J_2=0.15, ~J_0=0.15,~ J_p = 0.2$, and $\om = 3$. We
discover numerically that there are 50 Floquet edge modes; of these,
14 have FE very close to $+1$ and 36 have FE very close to $-1$.
Further, we discover that there are
Floquet modes on both zigzag and armchair edges. (An example of a
Floquet zigzag edge mode is shown in Fig.~\ref{fig08}). This is in contrast
to the time-independent version of the model discussed in Sec. III C which
has only 14 Majorana edge modes, all of which lie on the armchair edges.

The appearance of Floquet modes on both kinds of edges in this system can be
understood as follows. As discussed in Sec. III, the system with the
time-independent part of the Hamiltonian (i.e., with $J_3 = J_0$) lies in the
$A_x$ phase and therefore has Majorana modes only on armchair edges or at
corners as shown in Figs.~\ref{fig06}-\ref{fig07}. At the times $t=nT$,
Eq.~\eqref{ht1} shows that $J_3$ is infinitely large; if the couplings are
normalized to satisfy $J_1 + J_2 + J_3 = 1$, the system at these times
will lie at the top vertex $(0,0,1)$ in Fig.~\ref{fig03} and should therefore
have Majorana modes on zigzag edges. We therefore expect the kicked system to
have Majorana modes on both edges.

\subsection{Relation between bulk and edge modes}

We can understand the Floquet Majorana modes at the edges from the properties 
of the bulk modes as follows. For the infinite system with translation 
symmetry, the modes $(\hat a_{\vec k},\hat b_{\vec k})$ with different values 
of $\vec k$ decouple from
each other; hence we can study the Floquet operator $U_{\vec k} (T,0)$ for
each $\vec k$ separately. We then see from Eq.\ \eqref{hmom} that
\bea U_{\vec k} (T,0) &=& e^{-i2J_p \tau^y} ~e^{-iT (X_{\vec k} \tau^z ~+~
Y_{\vec k} \tau^y)}, \non \\
X_{\vec k} &=& 2 ~[J_1 \sin({\vec k} \cdot {\vec M}_1) ~-~ J_2 \sin({\vec k}
\cdot {\vec M}_2 )], \non \\
Y_{\vec k} &=& 2~[ J_0 ~+~ J_1 \cos({\vec k} \cdot {\vec M}_1 ~+~ J_2
\cos({\vec k} \cdot {\vec M}_2)]. \non \\
&& \label{bulkeq1} \eea
We will assume that $2J_p/\pi$ is not equal to an integer.
According to Ref.~\onlinecite{thakurathi}, a Majorana edge mode should appear
or disappear when $U_{\vec k}(T,0)$ has FE equal to $\pm 1$.
The structure of Eq.~\eqref{bulkeq1} implies that this will happen if
\beq J_1 \sin({\vec k} \cdot {\vec M}_1) - J_2 \sin({\vec k} \cdot
{\vec M}_2) ~=~ 0, \label{cond1} \eeq
and
\beq 2J_p + 2T\left[J_0 +J_1 \cos({\vec k} \cdot {\vec M}_1) +J_2
\cos({\vec k} \cdot {\vec M}_2) \right] ~=~ n\pi, \label{cond2} \eeq
where $n$ is an integer. [$U_{\vec k}(T,0)$ will have a FE equal to $+1$ ($-1$)
if $n$ is even (odd)]. We can use Eqs.~(\ref{cond1}-\ref{cond2}) to find the
critical values of $\om_{\vec k}$ where Majorana edge modes appear or
disappear. Since
\bea && ~~~[(J_1 \sin({\vec k} \cdot {\vec M}_1) ~-~ J_2 \sin({\vec k} \cdot
{\vec M}_2)]^2 \non \\
&& +~ [J_1 \cos({\vec k} \cdot {\vec M}_1) ~+~ J_2 \cos({\vec k} \cdot
{\vec M}_2)]^2 \non \\
&& =~ J_1^2 ~+~ J_2^2 ~+~ 2 J_1 J_2 \cos (k_x), \eea
Eq.~\eqref{cond1} implies that
\bea && J_1 \cos({\vec k} \cdot {\vec M}_1) +J_2 \cos({\vec k} \cdot
{\vec M}_2) \non \\
&& =~ \pm \sqrt{J_1^2 + J_2^2 + 2 J_1 J_2 \cos (k_x)}. \eea
Then Eq.~\eqref{cond2} implies that the critical values of $\om = 2\pi/T$
are given by
\beq \om_{\vec k} ~=~ \frac{4 \pi ~[~J_0 ~\pm~ \sqrt{J_1^2 + J_2^2 + 2 J_1
J_2 \cos (k_x)}~]}{n\pi ~-~ 2 J_p} \label{cond3} \eeq
which depends on a single momentum $k_x$.
For a system with a finite width bounded by infinitely long zigzag edges along
the $x$ direction, the momentum $k$ shown in Fig.~\ref{fig02} is equal to
$k_x$. We therefore have the prediction that for such a finite system, Floquet
Majorana modes should appear or disappear at the edges with a given value of
$k$ at frequencies which are given by Eq.~\eqref{cond3}. Further, the
Majorana mode should have a FE equal to $(-1)^n$.



This result can be generalized to systems in arbitrary dimensions. Let us
consider a $d$-dimensional system in which there are pairs of modes with
momenta $\vec k$ which are governed by a Hamiltonian of the form
\beq H'_{\vec k} ~=~ 2[\ep_{\vec k} \tau^y + \De_{\vec k} \tau^z], \eeq
where a component of $\ep_{\vec k}$ changes due to periodic kicks with a
frequency $\om$. Let $\ep_{1 \vec k}$ and $\ep_{0 \vec k}$ be the
values of $\ep_{\vec k}$ during and between the kicks. The Floquet
operator is then given by
\beq U_{\vec k} (T,0) ~=~ e^{-i2\ep_{1 \vec k} \tau^y} ~e^{-i2T (\De_{\vec k}
\tau^z ~+~ \ep_{0 \vec k} \tau^y)}. \eeq
If $2\ep_{1 \vec k}/\pi$ is not equal to an integer,
we can show that $U_{\vec k} (T,0)$ can have FE equal to $\pm 1$ only
if $\De_{\vec k} = 0$. Next, $\De_{\vec k} = 0$ will
generally define a $(d-1)$-dimensional hypersurface of the $d$-dimensional
Brillouin zone. Hence the frequency at which Majorana modes will appear or
disappear, $\omega_{\vec k}$, will depend on $d-1$ momenta and will
be determined by the conditions
\bea \De_{\vec k} &=& 0, \non \\
\om_{\vec k} &=& 4 \pi \ep_{0 \vec k}/(n \pi-2\ep_{1 \vec k}),
\label{condgen} \eea
where $n$ is an integer.
If we consider a system with a finite width which is bounded by two infinitely
large $(d-1)$-dimensional surfaces, there will generally be Floquet Majorana
modes on these surfaces which are parameterized by $d-1$ momenta.
Eq.~\eqref{condgen} will then determine the frequencies at which these modes
(with FE equal to $(-1)^n$) appear or disappear. For instance, the 2D
Kitaev model has $d=2$ so that the Floquet Majorana modes and the critical
frequencies $\om_{\vec k}$ depend on a single momentum, while the 1D Ising
model or Kitaev chain has $d=1$ so that the Majorana modes and the critical
frequencies are independent of any momentum~\cite{thakurathi}.

In the next subsection, we will use the above ideas to arrive at a
better understanding of the Floquet Majorana modes by mapping the Kitaev
honeycomb model to the 1D Kitaev chain where the Floquet problem has been
studied in detail earlier~\cite{thakurathi}.

\subsection{Mapping from the honeycomb model to a one-dimensional chain}

Consider a system which has a finite width in the $y$-direction (with zigzag
edges along the top and bottom as indicated in Fig.~\ref{fig02}) and is
infinitely long in the $x$-direction. The momentum $k$ along the $x$-axis is
a good quantum number. We now use the Heisenberg equations of motion
\bea \frac{d\hat a_m}{dt} &=& (J_1 e^{-ik/2} ~+~ J_2 e^{ik/2}) \hat b_m ~+~ 
J_3 \hat b_{m+1}, \non \\
\frac{d\hat b_m}{dt} &=& -(J_1 e^{ik/2} + J_2 e^{-ik/2}) \hat a_m - J_3 
\hat a_{m-1}, \label{eom2} \eea
for all $m \ge 1$, with the understanding that $\hat a_0 = 0$.

\begin{figure}[htb] \ig[width=2.in]{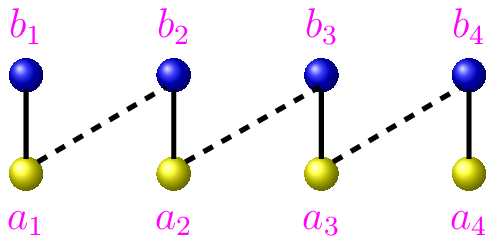} \\
\vspace*{.2cm}
\ig[width=1.5in]{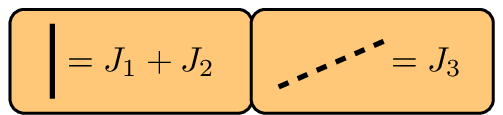}
\caption[]{(Color online) Mapping from the Kitaev honeycomb model to a
one-dimension chain for $k=0$, with the $J_1+J_2$ couplings shown as solid
lines and $J_3$ shown as dashed lines.} \label{fig09} \end{figure}

We first consider the case $k=0$. Then this problem converts in a
straightforward way to a special case of the 1D Kitaev chain (a system
of electrons with $p$-wave superconductivity)
with couplings as shown in Fig.~\ref{fig09}. This chain
is described by the Hamiltonian~\cite{thakurathi,gottardi}
\bea H &=& i \sum_{n=1}^\infty ~[ - J_x \hat b_{n+1} \hat a_n - J_y \hat b_n 
\hat a_{n+1} + \mu \hat b_n \hat a_n ]. \label{ham2} \eea
The Heisenberg equations of motion of the operators $\hat a_n$ and $\hat b_n$ 
in Eq.~\eqref{ham2} agree with Eqs.~\eqref{eom2} with $k=0$ if we set
\beq J_x ~=~ J_3/2, ~~J_y ~=~ 0, ~~{\rm and}~~ \mu ~=~ -(J_1 + J_2)/2.
\label{jxy} \eeq
Interestingly, this system is equivalent, by a Jordan-Wigner
transformation~\cite{lieb}, to an Ising model in a transverse
magnetic field described by the Hamiltonian
\beq H ~=~ - ~\sum_{n=1}^\infty ~[ J_x \si_n^x \si_{n+1}^x ~+~ \mu \si_n^z].
\label{ham3} \eeq

We now see that if $J_3$ is given periodic $\de$-function kicks in the
honeycomb model, it corresponds, for $k=0$, to a 1D model
in which the parameter $J_x$ is given periodic $\de$-function kicks without
changing the values of $\mu$ and $J_y$. This problem has been studied in
Ref.~\onlinecite{thakurathi}. It is known numerically (and analytically for
the special case $J_0 = 0$) that the $\de$-function kicks can produce Floquet
Majorana modes at the ends of the 1D system, which correspond to the zigzag
edges of the 2D model. In fact, we find numerically that one Floquet Majorana
mode appears at each of the zigzag edges (at the top and at the bottom of the
2D system) when the kicking frequency $\om$ is taken to be very large.

Next, we consider what happens if $k \ne 0$. In this case, we can
rewrite two of the parameters appearing in Eqs.~\eqref{eom2} as
\bea & & J_1 e^{\pm ik/2} ~+~ J_2 e^{\mp ik/2} ~=~ J_k e^{\pm i\phi_k}, \non \\
{\rm where} && J_k ~=~ \sqrt{J_1^2 ~+~ J_2^2 ~+~ 2 J_1 J_2 \cos k}.
\label{jjj} \eea
We can then show that the phase $\phi_k$ can be removed from Eqs.~\eqref{eom2}
by a unitary transformation; this unitary transformation is independent of
$J_3$ and is therefore not affected by the periodic kicks in $J_3$. We can
therefore study the problem just as in the case with $k=0$, except that the
parameter $\mu$ in Eq.~\eqref{jxy} is now given by $\mu_k = - J_k/2$. We thus
have a family of 1D problems which are labeled by the parameter $k$. For each
$k$, we look for Floquet edge modes. If we find such a mode, we can use the
idea discussed in Secs. III A and III B for the time-independent problem to
superpose the modes for the Floquet problems with $+k$ and $-k$ to obtain
a Majorana mode with Hermitian operators.

\begin{figure}[htb] \ig[width=2.8in]{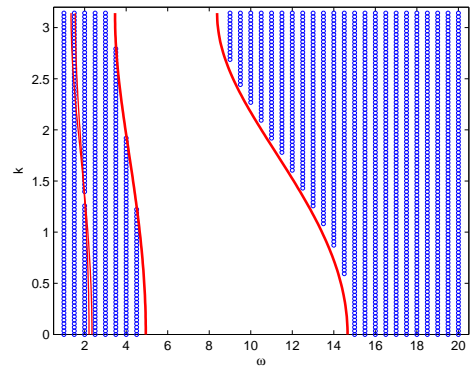}
\caption[]{(Color online) Blue regions in the $(\om,k)$ space where
Majorana states appear on the zigzag edges when the parameter $J_3$ is given
periodic $\de$-function kicks. The system being considered has a width of
$100$ sites, and $J_1=0.7, ~J_2=0.15, ~J_0=0.15$ and $J_p = 0.3$. The empty
region in the middle is bounded on the right and left by two solid red lines
which show the analytical results given in Eqs.~\eqref{cond6} and \eqref{cond7}
respectively. Two more red solid lines corresponding to $n=-1$ and 2 are
shown. They almost coincide with each other and appear within the blue
regions on the left; they cross near $\om=2$ and $k \simeq 1.3$.}
\label{fig10} \end{figure}

We have used the procedure described above to numerically find the region in
the space of $\om$ (from 1 to 20) and $k$ (from 0 to $\pi$) where Floquet
Majorana modes appear. Fig.~\ref{fig10} shows this for a system with a width
of $100$ sites (i.e., the index $m$ for $a_m$ and $b_m$ goes from 1 to 100),
with $J_1=0.70, ~J_2= 0.15, ~J_0=0.15$ and $J_p = 0.3$. If $\om$ is
sufficiently large, there is a Floquet Majorana mode with FE equal to $+1$.
As $\om$ is decreased, there is an empty region in which there are no Majorana
modes for any $k$. As $\om$ is decreased further, Majorana modes appear with
FE equal to $-1$. As explained in more detail below, the figure also shows
four red solid lines; two of these bound the empty region from the right and
left, while the other two almost coincide and lie within the blue region.
When $\om$ is decreased below the last two lines, the Majorana mode with FE
equal to $-1$ disappears and a mode with FE equal to $+1$ appears.

We now check how well these numerical results agree with the analysis given
in the previous subsection. According to Eqs.~\eqref{cond3} and \eqref{jjj},
Floquet edge modes with a given momentum $k$ and FE equal to $(-1)^n$
should appear or disappear when
\beq \om_k ~=~ \frac{4\pi ~[J_0 ~\pm~ J_k]}{n\pi ~-~ 2J_p}, \label{cond5} \eeq
where $n$ is an integer. As the kicking frequency $\om$ is decreased,
Eq.~\eqref{cond5} gives the red solid line on the right side of the empty
region
in Fig.~\ref{fig10} where a Floquet edge mode disappears with $n=0$, namely,
\beq \om_k ~=~ \frac{4 \pi ~[J_k ~-~ J_0]}{2 J_p}, \label{cond6} \eeq
and the red solid line on the left side of the empty region in Fig.~\ref{fig10}
where a Floquet edge mode appears with $n=1$, namely,
\beq \om_k ~=~ \frac{4 \pi ~[J_0 ~+~ J_k]}{\pi ~-~ 2 J_p}. \label{cond7} \eeq
In general, for $n \le 0$, we have a line given by
\beq \om_k ~=~ \frac{4 \pi ~[J_k ~-~ J_0]}{2 J_p - n \pi}, \label{cond8} \eeq
while for $n \ge 1$, we have a line given by
\beq \om_k ~=~ \frac{4\pi ~[J_0 ~+~ J_k]}{n\pi ~-~ 2J_p}, \label{cond9} \eeq
where Majorana modes appear or disappear. Fig.~\ref{fig10} also show the red
solid lines for $n=-1$ and 2. These appear within the blue regions;
they cross near $\om=2$ and $k \simeq 1.3$ where we see a small gap
indicating that there are no Majorana modes in that region. When
$\om$ is decreased below these two lines, the Majorana mode with
FE equal to $-1$ disappears and a mode with FE equal to $+1$
appears. When $\om$ is decreased even further, more modes start appearing
which correspond to $n > 2$ and $n < -1$.

\subsection{Periodic $\de$-function kicks in $J_1$ and $J_2$}

We have also studied what happens if we consider a finite system and apply
periodic kicks to $J_1$ or $J_2$, rather than to $J_3$. The time-independent
part of the Hamiltonian has $J_1 =0.7, ~J_2 = 0.15$ and $J_3 = 0.15$; such
a system lies in the $A_x$ phase and therefore only has edge states
on armchair edges. A periodic kick in $J_1$ means that at times
$t=nT$, the value of $J_1$ is infinitely larger than $J_2$ and $J_3$;
hence the system lies at the vertex $(1,0,0)$ in Fig.~\ref{fig03} which
also lies in the $A_x$ phase. Similarly, a periodic kick in $J_2$ means that
at $t=nT$, the value of $J_2$ is infinitely larger than $J_1$ and $J_3$;
the system then lies at the vertex $(0,1,0)$ in Fig.~\ref{fig03} which
lies in the $A_y$ phase. In both cases, we only expect edge states on
armchair edges.

For a system with $N_x \times N_y = 27 \times 14$ with a kick in $J_1$ or
$J_2$ with amplitude $J_p = 0.2$ and frequency $\om = 3$, we find numerically
that there are 14 Floquet Majorana modes, of which 12 are on the armchair
edges and 2 are at the corners. Thus the number and location of the Majorana
modes remain exactly the same as in the time-independent case discussed in
Sec. III C.

\section{Conclusions}
\label{diss}

In this work we have studied both equilibrium and Floquet edge modes of
the Kitaev model on a honeycomb lattice. One reason for studying the
Kitaev model is that it is the minimal model in two dimensions where
one can study edge states and derive a number of analytical results.
These results can be immediately applied to graphene for the following
reason. Graphene has ordinary fermion operators $c$ and $c^\dg$ which can 
be written in terms of two Majorana fermion operators $m_i$ at each site, 
as $c = (1/2) (m_1 + i m_2)$ and $c^\dg = (1/2) (m_1 - i m_2)$. The
Hamiltonian of graphene then turns out to be equivalent to two decoupled 
copies of the Kitaev Hamiltonian (four copies if we include the electron 
spin in graphene). This implies that if edge states appear in the Kitaev
model under some conditions, they must also appear in graphene under the
same conditions.

We have discussed the known 
analytical solutions for the equilibrium zero energy modes localized at
both zigzag and armchair edges. These solutions lead to a phase diagram for
the presence or absence of these modes and the possible values of their
momentum along the edge. These states, in
contrast to the bulk modes, have wave function weight on only one of the
sublattices. We have pointed out that this property provides a way of
spatially separating the sublattice constituents of a Majorana fermion and
have discussed this phenomenon in the context of standard electron
fractionalization found, for example, in polyacetylene and edges of
unconventional superconductors.

Next we have studied the Floquet edge modes which appear in the
Kitaev model when the coupling on the bonds perpendicular to the
edge is varied in time as periodic $\de$-function kicks. Using a
relation between the bulk and edge modes we have found a generic
condition on the drive frequency of a $d$-dimensional integrable model 
which needs to be satisfied for the appearance or disappearance of these 
edge modes. We have verified this generic condition in the Kitaev model 
with a finite width by mapping it to a one-dimensional model of electrons 
with $p$-wave superconductivity (or an Ising chain in a transverse 
magnetic field). We have shown that the $\de$-function kicks can lead to 
a large number of Floquet edge modes, and that these modes can appear on 
certain edges even when there are no equilibrium Majorana modes on those 
edges. Finally, we have supplemented our analytical calculations with 
numerical analysis for finite-sized systems which confirms the above 
prediction for the drive frequencies. In the context of Floquet modes, 
our numerics shows that edge modes can appear both at the edges and the 
corners of a finite sample as the drive frequency is varied.

We summarize our most important results as follows. \\
\noi (i) Periodic driving of some of the couplings of the Kitaev model
can give rise to edge states in certain regimes of couplings where
the time-independent part of the Hamiltonian has no edge states. \\
\noi (ii) The driving frequencies at which Majorana edge modes appear or
disappear in a two-dimensional system can be completely understood
by mapping it to a one-dimensional system in which the edge momentum
appears as one of the parameters of the model. 

There are proposals for realizing the Kitaev model in systems of cold atoms
trapped in optical lattices~\cite{optical1,optical2,zhang,han}. It may
therefore be possible to look for states localized at the edges of such
systems, both at equilibrium and in the presence of periodic driving. In the
latter case, it would be necessary to consider the effects of random noise
and various relaxation mechanisms which may be present in the
system~\cite{lind1,kita2,thakurathi}.

\vspace*{.4cm}
\section*{Acknowledgments}
For financial support, M.T. thanks CSIR, India and D.S. thanks DST, India
for Project No. SR/S2/JCB-44/2010. The authors thank R. Shankar for
stimulating discussions on related topics.

\end{document}